\documentclass[aps,prl,twocolumn,showpacs]{revtex4}
\usepackage{graphics}
\usepackage{dcolumn}
\usepackage{bm}

\begin{document}

\title{New CP-violation and preferred-frame tests with polarized electrons}
\author{B. R. Heckel}
\author{C. E. Cramer}
\author{T. S. Cook}
\author{E. G. Adelberger}
\author{S. Schlamminger}
\author{U. Schmidt}\altaffiliation[Current address: ]{Physicalisches Institute, Universit\"at Heidelberg, Philosophenweg 12, D-69120}

\affiliation{Center for Experimental Nuclear Physics and Astrophysics, Box 354290,
University of
Washington, Seattle, WA 98195-4290}
\date{\today}

\begin{abstract}
We used a torsion pendulum containing $\sim 9 \times 10^{22}$ polarized electrons to 
search for CP-violating interactions between the pendulum's electrons and unpolarized matter in the laboratory's surroundings or the sun, 
and to test for preferred-frame effects
that would precess the electrons about a direction fixed in inertial space. We find $|g_{\rm P}^e g_{\rm S}^N|/(\hbar c)< 1.7 \times 10^{-36}$ and $|g_{\rm A}^e g_{\rm V}^N|/(\hbar c) < 4.8 \times 10^{-56}$ for $\lambda > 1$AU.  Our preferred-frame constraints, interpreted in the Kosteleck\'y framework, set an upper limit on the  
parameter $|\bm{\tilde {b}}^e| \leq 5.0 \times 10^{-21}$~eV that should be compared to the benchmark value $m_e^2/M_{\rm Planck}= 2 \times 10^{-17}$~eV. 
\pacs{11.30.Cp,12.20.Fv}
\end{abstract}
\maketitle

This Letter reports constraints on proposed new spin-coupled interactions using a torsion pendulum containing $\sim 9 \times 10^{22}$ polarized electrons. Several lines of speculation motivated our work. We were motivated to consider preferred-frame effects because the cosmic microwave background does establish a such a frame. Kosteleck\'y and coworkers\cite{co:97} have discussed an unusual cosmic preferred-frame effect where vector and axial-vector fields could have been spontaneously generated in the early universe and then been inflated to enormous extents. They emphasize that these fields would provide a
mechanism for CPT and Lorentz violation. 
Because the fields invalidate the Pauli-Luders theorem,
one can construct a field theory with CPT- and
Lorentz-violating effects (the
Standard-Model Extension) and quantify the
sensitivity of various CPT and preferred-frame tests. 
One manifestion of such fields would be an anomalous torque on a spinning electron\cite{bl:00} arising from a potential 
\begin{equation}
V_e=\bm{\sigma}_e \cdot \bm{\tilde{b}}^e~,
\label{eq: BK}
\end{equation}
where $\bm{\tilde{b}}^e$
is the product of the presumed cosmic axial-vector field and its coupling to electrons. 

Spin-dependent forces are also generically produced by the exchange of pseudoscalar particles. Moody and Wilczek\cite{mo:84} discussed the forces produced by the exchange of low-mass, spin-0 particles and
pointed out that particles containing CP-violating $J^{\pi}=0^+$ and $J^{\pi}=0^-$ admixtures would produce a macroscopic, CP-violating ``monopole-dipole'' interaction between a polarized electron and an unpolarized atom with mass and charge numbers $A$ and $Z$ 
\begin{equation}
V_{e A}(r)= g_{\rm P}^e g_{\rm S}^A  \frac{\hbar}{8 \pi m_e c}{\bm{\sigma}}_e \cdot \left[ {\bm{\hat r}} \left( \frac{1}{r \lambda}+\frac{1}{r^2}\right)e^{-r/\lambda}\right]~, 
\label{eq: MW}
\end{equation} 
where $m_{\phi}=\hbar/(\lambda c)$ is the mass of the hypothetical spin-0 particle, $g_{\rm P}$ and $ g_{\rm S}$ are its pseudoscalar and scalar couplings, and $g_{\rm S}^A= Z(g_{\rm S}^e+g_{\rm S}^p) + (A-Z) g_{\rm S}^n$. For simplicity, we assume below that $g_{\rm S}^p=g_{\rm S}^n=g_{\rm s}^N$ and $g_{\rm S}^e=0$ so that $g_{\rm S}^A=A g_{\rm S}^N$; contraints for other choices of the scalar couplings can be readily obtained by scaling our limits. 

Recently Dobrescu and Mocioiu\cite{do:06} classified the kinds of potentials that might arise from exchange of low-mass bosons, constrained only by rotational and translational invariance. We are sensitive to 3 of their potentials; in addition to a potential equivalent to Eq.~\ref{eq: MW}, we probe two potentials that we write as   
\begin{equation}
V_{e N}(r)=\bm{\sigma}_e \cdot \left[A_{\perp}\frac{\hbar}{c}\frac{(\bm{\tilde{v} \times {\hat r}}
)}{m_e}\left( \frac{1}{r \lambda}+\frac{1}{r^2}\right) + 
A_v \frac{\bm{\tilde{v}}}{r}
\right]e^{-r/\lambda}~,
\label{eq: BD}
\end{equation}
where $\bm{\tilde {v}}$ is the relative velocity in units of $c$. Both terms may be generated by one-boson exchange in Lorentz-invariant theories.
The parity-conserving $A_{\perp}$ term can arise from
scalar or vector boson exchange, while the parity-violating $A_v$ term
can be induced by vector bosons that
have both vector and axial couplings to electrons or nucleons with $A_v=g_{A}^e g_{V}^N/(4 \pi)$.

Our work substantially improves upon the upper limits on 
$\bm{\sigma}_e \cdot \bm{\tilde{b}}^e$ interactions presented in
Ref.~\cite{ho:03}, on
$\bm{\sigma}_e \cdot \bm{r}$ interactions 
in Refs.~\cite{wi:91,yo:96,ni:99}, 
, and we obtain new constraints on the terms in Eq.~\ref{eq: BD}.
%
%
\begin{figure}[t]
\hfil\scalebox{.33}{\includegraphics*[0.9in,2.6in][7.8in,10.0in]{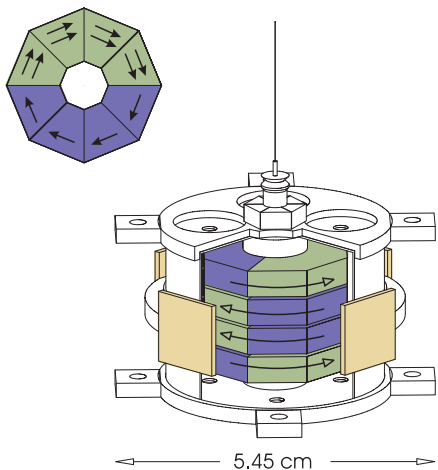}}\hfil
\caption{[Color online] Scale drawing of the spin pendulum. The light green and darker blue volumes are Alnico and Sm$\,$Co$_5$, respectively. Upper left: top view of a single ``puck''; the spin moment points to the right. Lower right: the assembled pendulum with the magnetic shield shown cut away to reveal the 4 pucks inside. Two of the 4 mirrors (light gold) used to monitor the pendulum twist are prominent. Arrows with filled heads show the relative densities and directions of the electron spins, open-headed arrows show the directions of ${\bm B}$. The 8 tabs on the shield held small screws that we used to tune out the pendulum's residual $Q_{21}$ and $Q_{22}$ moments.}
\label{fig: spin pendulum}
\end{figure}

The heart of our apparatus is a spin pendulum, shown in Fig.~\ref{fig: spin pendulum}, that contains a substantial number of polarized electrons while having a negligible external magnetic moment and high gravitational symmetry. 
The spin pendulum is constructed from 4 octagonal ``pucks''. One side of each puck is Alnico (a conventional ``soft'' ferromagnet in which the magnetic field is created almost entirely by electron spins) and the other side from Sm$\,$Co$_5$ (a ``hard'' rare-earth magnet in which the orbital magnetic moment of the electrons in the Sm$^{3+}$ ion \cite{ko:97,ti:99,gi:79} nearly cancels their spin moment).After each puck was assembled, we 
magnetized the Alnico to the same degree as the Sm$\,$Co$_{5}$
by sending appropriate
current pulses through coils temporarily wound around the pucks.
By stacking 4 such pucks as shown in Fig.~\ref{fig: spin pendulum}, we placed the effective center of the spin dipole in the middle of the pendulum, reduced systematic magnetic-flux leakage, averaged out the small density differences between Alnico and Sm$\,$Co$_{5}$, and cancelled any composition dipole that would have made us sensitive to violation of the weak Equivalence Principle.

We estimated the net spin of the pendulum using results from circularly-polarized X-ray Compton scattering\cite{ko:97} and polarized-neutron scattering\cite{gi:79,ti:99} studies of  Sm$\,$Co$_{5}$. The X-ray study  found that at room-temperature the ratio of Co to Sm spin moments is $R=-0.23\pm0.04$, while the neutron work showed that the Sm magnetic moment is very small, 0.04$\mu_{B}$ vs. 7.8$\mu_{B}$ per formula unit for the Co. Therefore the magnetization of Sm$\,$Co$_{5}$ is due almost entirely to the Co, so that the Co and Alnico contributions to the net spin moment of our pendulum cancel. The net moment arises essentially entirely from the Sm spins. Then the number of polarized spins in our pendulum is 
\begin{equation}
N_{\rm p}=\frac{B_0 R}{\mu_0 \mu_B} V \eta =6\times10^{22}~,
\label{eq: Npol}
\end{equation}
where $B_0$ is the magnetic field inside a puck, $\eta=0.65$ accounts for its octagonal shape and $V=9.81$ cm$^3$ is the total volume of the pucks. 
We measured $B_0$, the field inside identical Sm$\,$Co$_{5}$ elements arranged in straight line,
using an induction coil and found $B_0=9.5$ kG which agreed with the supplier's specification.  The Sm ion wavefunctions deduced from neutron scattering\cite{gi:79} predict a
room-temperature Sm spin moment for Sm in Sm$\,$Co$_{5}$ of $-3.59\mu_{B}$. This is equivalent to $R \approx-0.44$ and implies $N_{\rm p}=11\times10^{22}$. We assume, in deriving our constraints below, that $N_{\rm p}$ is equally likely to have any value between $6\times 10^{22}$ and $11\times 10^{22}$.

Our pendulum was suspended by a $28\mu$m diameter, 75 cm long tungsten fiber inside a uniformly rotating torsion balance that is an upgraded version of the instrument described previously\cite{ba:99}. The pendulum's free-oscillation frequency, $f_0=2.570$ mHz, together with its calculated rotational inertia, determined the fiber's torsional constant $\kappa=0.118$ dyne-cm/radian. The main improvement in our turntable was a ``feet-back'' system that kept its rotation axis vertical to better than 10 nradians, continuously correcting for the varying tilt of the laboratory floor and imperfections in the turntable bearing by
controlling the temperature (and thereby the length) of feet that support the turntable. In addition, we improved the co-rotating mu-metal magnetic shielding.
%
%
\begin{figure}[b]
\hfil\scalebox{.9}{\includegraphics*[40pt,31pt][313pt,291pt]{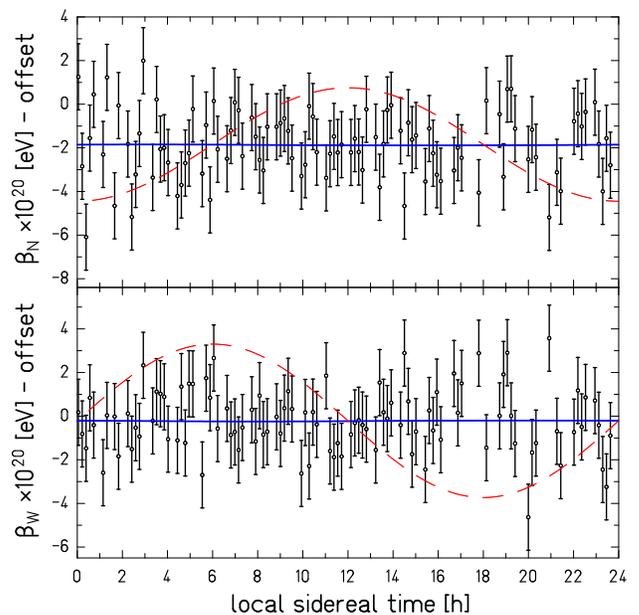}}\hfil
\caption{Data from a set of runs at $\phi_{\rm d}=22.5$ deg. The turntable offset (determined from data at all 4 $\phi_{\rm d}$ values shown in Fig.~\ref{fig: 4 phis}) is subtracted from the vertical axes. The dashed curves show the signal from a hypothetical $\bm{\tilde{b}}^e=(5 \times 10^{-20}~{\rm eV})\bm{\hat{x}}$ which gives out-of-phase sine waves in $\beta_N$ and $\beta_W$. The solid curves show the best sidereal fit, which yields $\tilde{b}_x=(-0.2 \pm 1.9)\times 10^{-21}$ eV, $\tilde{b}_y=(-0.1 \pm 1.9)\times 10^{-21}$ eV. 
The data set spanned a duration of 118 h.}
\label{fig: typical data}
\end{figure}

We recorded the pendulum twist angle as a function of the angle of the turntable, and converted these to torques, as described in Ref.~\cite{su:94}. Data were accumulated over a span of 13 months and divided into ``cuts'' each of which spanned no more than 3800 s. From each cut, we extracted the component of the twist signal that varied harmonically at the turntable rotation frequency $f$; over the course of this experiment $f$ was set at values between $3f_0/29$ and $3f_0/20$ . Data were taken with 4 equally-spaced angles
$\phi_{\rm p}$ of the {\em pendulum} within the rotating apparatus. Averaging these 4 results  cancelled the effects of any steady irregularities in the turntable rotation rate. We did not know the orientation $\varphi$ of the {\em spin dipole} inside the pendulum until our extraction of the torque signals was complete. Only then did we remove the shield, find $\varphi$ so we could learn the orientations of the spin dipole with respect to the turntable, $\phi_{\rm d}=\phi_{\rm p}+\varphi$.
With that information, we could convert the twist signals for each cut into the
North and West components of $\bm{\beta}$, where the energy $E$ of spin dipole $\bm{\mu}_{\rm s}$ was $E=-N_{\rm p}\,\bm{\hat\mu}_{\rm s} \cdot \bm{\beta}$.

We first analysed our data for signals modulated at solar or sidereal periods,
using astronomical formulae given by Meeus\cite{me:98}.
We constrained $\bm{\tilde b}^e$ in Eq.~{\ref{eq: BK} by fitting for
signals corresponding to a $\bm{\beta}$ fixed 
in inertial space using methods similar to those described in Ref.~\cite{su:94}. Figure~\ref{fig: typical data} shows typical data set.
In this case we employed a rectilinear coordinate system where $\bm{z}$ is parallel to the earth's rotation axis, $\bm{x}$ lies along the vernal equinox and $\bm{y} = \bm{z \times x}$.
Both $\tilde{b}^e_x$ and $\tilde{b}^e_y$ generate $\beta_{\rm N}$ and  $\beta_{\rm W}$ signals that are modulated at the sidereal rate, while $\tilde{b}^e_z$ produces a steady $\beta_{\rm N}$ signal. The sidereal modulation eliminates many systematic effects that are fixed in the lab; as a consequence our bounds on  $\tilde{b}^e_x$ and $\tilde{b}^e_y$, shown in Table~\ref{tab: Kost params} are tighter than those on $\tilde{b}^e_z$ which are based on the lab-fixed limits discussed below. 

\begin{table}[h]
\caption{$1\sigma$ constraints on the Kosteleck\'y $\bm{\tilde{b}}^e$ parameters from  
our work and from Hou et al.\cite{ho:03} Units
are $10^{-22}$ eV.}
\begin{ruledtabular}
\begin{tabular}{ccc}
parameter  & this work    &  Hou et al. \\
\colrule
$\tilde{b}_{\rm x}^e$  &   $+0.1\pm 2.4$   &   $-108\pm 112$ \\
$\tilde{b}_{\rm y}^e$  &   $-1.7\pm 2.5$  &   $-5\pm 156$ \\
$\tilde{b}_{\rm z}^e$  &   $-29\pm 39$     &   $107\pm 2610$ \\
\end{tabular}
\end{ruledtabular}
\label{tab: Kost params}
\end{table}

We constrained the terms in Eq.~\ref{eq: BD} and the long-range limit of Eq.~\ref{eq: MW} by considering interactions between our spin pendulum and the sun. 
Because of the 23.45 degree inclination of the earth's rotation axis, these
torques have components modulated with a 24 hour period as well as 
annual modulations. Our constraints, shown in Table~\ref{tab: Dobr params}, are based on the modulated signals, with the individual runs weighted by the inverse squares of their errors.

\begin{table}[hb]
\caption{$1\sigma$ constraints from interactions with the Sun. These values assume $\lambda > 1 {\rm AU}$.}
\begin{ruledtabular}
\begin{tabular}{lc}
parameter  & constraint  \\
\colrule
$g_{\rm P}^e g_{\rm S}^N/(\hbar c)$  &   $(-0.4\pm 1.6)
 \times 10^{-36}$  \\
${A_\perp}/(\hbar c)$  &   $(-2.4\pm 6.4) \times 10^{-34}$  \\
$A_v/(\hbar c)=g_{\rm A}^e g_{\rm V}^N/(4 \pi \hbar c)$  &   $(+3.0\pm 1.7) \times 10^{-57}$  
\end{tabular}
\end{ruledtabular}
\label{tab: Dobr params}
\end{table}
The dominant sources of systematic errors are possible daily variations of the tilt, temperature or vibration of the apparatus, and of external gravity gradients or magnetic fields. 
We measured the sensitivity of our apparatus to each source by applying a known, magnified change in that source. 
We deduced a systematic error by multiplying the sensitivity by the daily signal recorded by sensors that monitored each source. No significant systematic error was found. The experimental errors in Tables \ref{tab: Kost params} and \ref{tab: Dobr params} are the quadrature sum of statistical uncertainties and upper limits on systematic errors that never exceeded the statistical uncertainty.
%
%
\begin{figure}[b]
\hfil\scalebox{.75}{\includegraphics*[0.68in,0.4in][4.4in,4.1in]{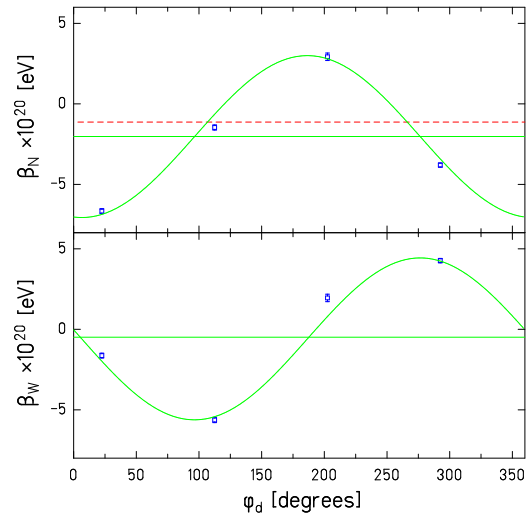}}\hfil
\caption{Extraction of lab-fixed signals from one of the complete sets of 4 $\phi_{\rm d}$ values. The sine and cosine waves result from turntable-fixed effects while solid horizontal
lines are the lab-fixed signals; the gyrocompass effect is shown by the dashed line. The data in Fig.~\ref{fig: typical data} yielded the $\phi_{\rm d}=22.5^{\circ}$ points.}
\label{fig: 4 phis}
\end{figure}

Finally, we analyzed our $\beta_{\rm N}$ and $\beta_{\rm W}$ signals for torques fixed in the lab frame by  comparing the signals observed for 4 equally-spaced angles, $\phi_{\rm d}$, of 
$\bm{\mu}_{\rm s}$ within the rotating apparatus (see Fig.~\ref{fig: 4 phis} and Table~\ref{tab: 4 phitops}). 
In Fig.~\ref{fig: 4 phis}, signals from a steady anomaly in the turntable rotation rate are  sinusoidal functions of $\phi_{\rm d}$, while a torque that coupled to $\bm{\mu}_{\rm s}$ would show up as non-zero averages of the 4 $\beta_{\rm N}$ and $\beta_{\rm W}$ values. We combine 11 such measurements in Table~\ref{tab: 4 phitops}. The individual entries in Table~\ref{tab: 4 phitops} are corrected for the dominant systematic errors: small, residual couplings to lab-fixed gravity gradients and magnetic fields.
Sensitivities to these couplings were found by applying known, large gravity gradients and by reversing the current in the Helmholtz coils that surround the apparatus. Corrections were obtained by multiplying the sensitivities by the the measured gravity gradients and magnetic fields present during normal data collection. The error quoted in Table~\ref{tab: 4 phitops} is based on the scatter of the 11 measurements and includes the uncertainty in $N_{\rm p}$. This scatter is larger than our statistical uncertainties and is still under investigation. 

Because the pendulum's magnetic flux was confined entirely within the pucks, the total intrinsic angular momentum of the pendulum was $J_3=-S_3$, where $S_3=N_{\rm p} \hbar/2$ is the pendulum's net spin.
The earth's rotation $\bm{\Omega_{\oplus}}$ acting on $\bm{J}$ of the electrons produced a steady torque along the suspension fiber $|\bm{\Omega_{\oplus} \times J \cdot \hat{n}}|$ ($\bm{\hat{n}}$ is the local vertical) equivalent to a
small negative (because $\bm{J}=-\bm{S}$) value $\beta_{\rm N}=-1.61\times 10^{-20}$~eV. Table \ref{tab: 4 phitops} shows that this gyrocompass effect was detected; it was subtracted from the measured $\beta_{\rm N}$ to
constrain $\tilde{b}_{\rm z}^e$ in Eq.~\ref{eq: BK}  and $g_{\rm P}^e g_{\rm S}^A$ in 
Eq.~\ref{eq: MW}. 
The latter constraints  depend on the horizontal component of the term in square brackets in  Eq.~\ref{eq: MW}. We integrated this term over the local mass distribution consisting of the significant masses in the laboratory and its surrounding topography as described in Ref.~\cite{ad:90}. This integral is, within a constant, identical to the integral $\bm{J}_{\perp}(\lambda)$ defined and evaluated in Ref.~\cite{ad:90}.
Figure~\ref{fig: constraints} shows our constraints on the product $g_{\rm P}^e g_{\rm S}$  in Eq.~\ref{eq: MW}.
\begin{table}[t]
\caption{Lab-fixed signals
, $\beta_N$ and  $\beta_W$,
 extracted from 11 complete data sets
, each containing measurements at all 4 values of $\phi_{\rm d}$.
Signals from each data set are corrected for measured gravity-gradient and magnetic couplings that were less than $(0.15 \pm 0.02) \times 10^{-20}$ eV and $(0.47 \pm 0.09) \times 10^{-20}$ eV, respectively. 
Errors in the net result are the larger of the uncertainties in the two averages.}
\begin{ruledtabular}
\begin{tabular}{lccr}
set  & dates &  $\beta_N\times 10^{20}$    &   $\beta_W\times 10^{20}$ \\
   & mo/day/yr   &  (eV)   &   (eV)  \\
\colrule
~1   & 08/19/04~~to~~09/10/04~~~  &   $-2.59 $ &   $-0.61 $  \\
~2   & 11/18/04~~to~~12/09/04~~~  &   $-2.18 $ &   $-0.19 $  \\
~3   & 12/21/04~~to~~01/06/05~~~  &   $-1.89 $ &   $-0.74 $  \\
~4   & 01/13/05~~to~~02/03/05~~~  &   $-1.84 $ &   $-0.28 $  \\
~5   & 05/27/05~~to~~06/11/05~~~  &   $-0.73 $ &   $-0.51 $  \\
~6   & 06/20/05~~to~~06/24/05~~~  &   $-0.93 $ &   $+0.11 $  \\
~7   & 06/26/05~~to~~06/30/05~~~  &   $-0.52 $ &   $-0.31 $  \\
~8   & 08/06/05~~to~~08/14/05~~~  &   $-0.37 $ &   $-0.49 $  \\
~9   & 12/25/05~~to~~12/29/05~~~  &   $-0.59 $ &   $+0.44 $  \\
~10  & 12/29/05~~to~~01/04/06~~~  &   $-0.70 $ &   $+0.23 $  \\
~11  & 01/08/06~~to~~01/12/06~~~  &   $-0.68 $ &   $-0.05 $  \\
\colrule
 & average $\pm$ uncertainty     &   $-1.19\pm 0.34$ &   $-0.23\pm 0.13$  \\
\colrule
 & gyro effect  & $-1.61$ & 0 \\
\colrule
 & net result               & $+0.42\pm 0.34$ &  $-0.23\pm 0.34$ \\
\end{tabular}
\end{ruledtabular}
\label{tab: 4 phitops}
\end{table}
%
%
\begin{figure}[t]
\hfil\scalebox{0.7}{\includegraphics*[51pt,28pt][394pt,291pt]{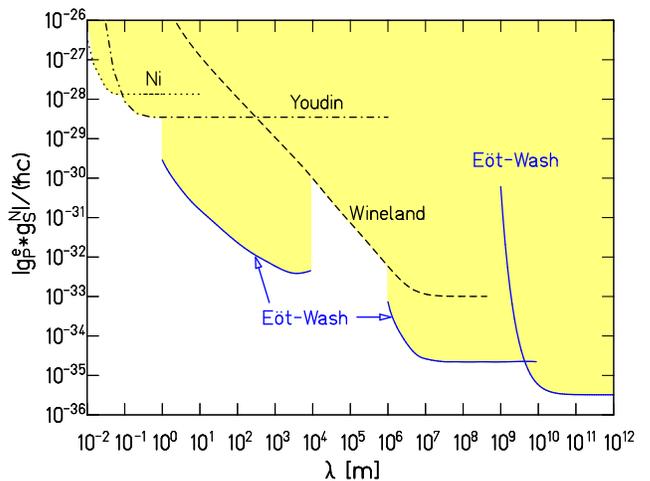}}\hfil
\caption{[Color online]Upper limits on $|g_{\rm P}^e g_{\rm S}^N|/(\hbar c)$ as a function of interaction range $\lambda$; the shaded region is excluded at  2$\sigma$. Our results and previous work by Youdin et al.\cite{yo:96}, Ni et al.\cite{ni:99} and Wineland et al.\cite{wi:91} are indicated by solid, dash-dotted, dotted and dashed lines, respectively.  
Our work does not provide constraints for 10 km $< \lambda < 10^3$ km because integration over the terrestrial surrounding is not reliable in this regime (see Ref.~\cite{ad:90}).} 
\label{fig: constraints}
\end{figure}

In summary, we have shown that a torsion balance fitted with a spin pendulum can achieve
a constraint of $\sim 10^{-21}$ eV on the energy required to flip an electron spin about directions fixed in inertial space. This is comparable to the electrostatic energy of two electrons separated by 10 AU. We then use these and related
constraints to set sensitive limits on preferred-frame, CP-violating, and velocity-dependent P-violating interactions of electrons. Constraints on preferred-frame effects involving protons and neutrons are given in Refs.~\cite{ph:01} and \cite{ca:04}, and on CP-violating electron-neutron interactions in Ref.~\cite{wi:91}. 

Michael Harris and Stefan Bae\ss ler  developed earlier versions of this apparatus and provided us with essential experience that made this work possible. 
Jens Gundlach and CD Hoyle contributed helpful discussions about this experiment. We thank Alan Kosteleck\'y and Bogdan Dobrescu for inspiring conversations, and Tom Murphy, Jr for advice on the astronomical calculations. This work was 
supported by NSF Grant PHY0355012 and by the DOE Office of Science. CEC is grateful for an NSF Fellowship.

\end{document}